# Ideal fully spin-polarized type-II nodal line state in half-metals $X_2YZ_4$ (X=K, Cs, Rb, Y=Cr, Cu, Z=Cl, F)

Tingli He,[a,b] Xiaoming Zhang,[a,b,*] Lirong Wang,[a,b] Ying Liu,[a,b] Xuefang Dai,[a,b] Liying Wang,[c] and Guodong Liu[a,b,†]

[a]State Key Laboratory of Reliability and Intelligence of Electrical Equipment, Hebei University of Technology, Tianjin 300130, China.

[b]School of Materials Science and Engineering, Hebei University of Technology, Tianjin 300130, China.

[c]Tianjin Key Laboratory of Low Dimensional Materials Physics and Preparation Technology, School of science, Tianjin University, Tianjin 300354, People's Republic of China.

* Correspondence: zhangxiaoming87@hebut.edu.cn; gdliu1978@126.com

**Abstract**

Lorentz-violating type-II nodal lines exhibit attracting physical properties and have been hot discussed currently. However, their investigations have been mostly limited in nonmagnetic system because of lacking ideal spin-polarized candidates with clean type-II nodal line states. Here, for the first time, we report the family of $X_2YZ_4$ (X=K, Cs, Rb, Y=Cr, Cu, Z=Cl, F) compounds are such ideal candidate materials by using the member of $K_2CuF_4$ as an example. We show the material is a ferromagnetic half-metal with weak anisotropy, which host fully spin-polarized conducting electrons. In the conducting spin channel, the band crossing form a pair of type-II nodal lines, protected by mirror symmetry. These type-II nodal lines are different with former proposed examples because they have a 100% spin polarization. In addition, we also show the material can realize switchable topological states, which can be easily controlled by external magnetic field. It is noticed that, the material: i) is stable and can be synthesized in experiments; ii) has clear magnetic structure; and iii) manifests clean type-II nodal line state and clear drumhead surface states. Therefore, the proposed $X_2YZ_4$ compounds are expected to be an excellent platform to investigate the novel physical properties of both type-II nodal line states with complete spin polarization.

## 1. Introduction

Recently, the interaction between band topology and crystalline symmetry has attracted considerable attention in materials science and condensed matter physics [1-6]. In addition to topological insulators, topological semimetals and metals, originating from the non-trivial band crossings near the Fermi energy, have been intensively investigated [7-11]. Notably, the protected band degeneracies of electronic structure by non-trivial band topology are of particular interest, leading to the discovery of various topological quasi-particle fermions, such as Majorana fermions [12-14], Weyl fermions [15-19], Dirac fermions [20,21], line nodes fermions [22,23], hourglass fermions [24,25], and three-component fermions [26]. Nodal line semimetals (NLSMs) represent a new type of topological semimetal phase different from Weyl and Dirac semimetals which feature closed loops or open curves of band degeneracy in the Brillouin zone (BZ). As with the classification of type-I and type-II Weyl semimetals, these two types of NLSMs are classified based on the tilting degree of bands [27]. The slopes of the two crossing bands of type-I NLSMs have opposite signs, while the slopes of the two energy bands of type-II NLSMs have the same sign along at least in one direction of the BZ. In the type-II nodal line semimetal, the point of each nodal line is strongly inclined in a lateral direction as in the type-II Weyl semimetal. Compared with the conventional type-I node line semimetals, there are significant differences in the type-II ones, such as the squeezing or collapse of the Landau levels [28-29], the Klein tunneling [30,31], and the magnetic breakdown in momentum space when a magnetic field is applied to an over-tilted Weyl node [32,33].

Until now, type-II nodal lines have received extensive attention and been investigated in several non-magnetic materials. In 2017, Li *et al.* proposed the concept of type-II node line and the first realistic possible material namely $K_4P_3$ based on theoretical calculations [27]. After that, Zhang *et al.* theoretically proposed $Mg_3Bi_2$ to be an ideal type-II NLSM with clean electronic structure and surface states [33], which has been confirmed by later experiments [34]. In addition to three-dimensional (3D) materials, type-II nodal lines are also proposed in peculiar two-dimensional (2D) material, such as TiBX (X = Cl, Br, I) monolayers [35].

In NLSMs, nodal lines can be protected related to different symmetries, including the coexistence of inversion and time-reversal symmetries [36,37], the mirror symmetry [38] and specific nonsymmorphic symmetries [39,40]. Recently, NLSMs in magnetic materials without the time-reversal symmetry have spurred a deal of research interest in exploring exotic phenomena, such as tunable nodal points [41] and anomalous Hall effect [42,43]. Especially, some ferromagnetic materials have been proposed to show type-II nodal lines such as 2D InC [44] and ScCl sheet [45]. In these examples, the bands from both spin channels coexist near the Fermi level, thus the extinctions from nodal line electrons are not completely spin polarized. Very recently, fully spin-polarized type-II nodal line is proposed in the ferromagnetic insulating phase $Sr_2SrOsO_6$ at peculiar condition [46], with the coexistence with multi-Weyl nodes near the Fermi level. This example has shown the feasibility of realizing fully spin-polarized type-II nodal line in realistic materials, but it still faces crucial issues. First, the material is an insulating phase and the electronic state for the type-II nodal line is difficult to contribute to the conducting behavior. Second, the type-II nodal line coexists with multi-Weyl nodes, then how to distinguish the nodal line fermion from other components could be a challenging task. Therefore, exploring ideal candidate materials which show fully spin-polarized conducting electrons and clean type-II nodal line states is still in urgent need for experimental detections and future applications.

In the work, we propose a family of quasi-two-dimensional compounds namely $X_2YZ_4$ (X = K, Cs, Rb, Y = Cr, Cu, Z=Cl, F) are such ideal materials with fully spin-polarized conducting electrons and clean type-II nodal line states based on first principles calculations. Our calculations suggest these materials naturally show the ferromagnetic ordering, being consistent with the existing experimental results. Taking $K_2CuF_4$ compound as an example, we find that it exhibits the band structure of a half metal, with a pair of type-II nodal lines in the spin-down channel. The nodal lines locate in the (110) plane, protected by the mirror symmetry $M_{110}$. It is worth noticing that, the type-II nodal line fermions are fully spin-polarized, which are previous ones proposed in nonmagnetic and conventional magnetic systems. If SOC is included, the symmetry of the system depends on the magnetization direction. In the most easily magnetized [001] direction, the type-II nodal lines will degenerate to two pairs of type-II Weyl nodes. However, the nodal lines can be nicely preserved if the magnetization along the [110] direction. According to former experiments and our calculations, $K_2CuF_4$ show weak anisotropy. Therefore, the material can realize switchable Weyl and nodal lines states, as easily controlled by external magnetic field. We also show that similar characters are also applied in other member of $X_2YZ_4$ compounds. Considering that these materials show fully spin-polarized conducting electrons and clean type-II nodal line states, they are an ideal platform to explore the type-II nodal line states with high spin polarization.

## 2. Crystal structure and magnetic properties

The crystal structure of $X_2YZ_4$ (X = K, Cs, Rb, Y = Cr, Cu, Z=Cl, F) compounds within the I4/*mmm* (No. 139) space group has been well characterized experimentally more than three decades ago [47-49]. These compounds have similar crystal structure, thus in the following we use the case of $K_2CuF_4$ as an example. In the crystal structure, each $Cu^{2+}$ ion is surrounded by six $Cl^-$ ions and forms an octahedron structure, as shown in Fig 1(a). Along the *c* direction, four $Cl^-$ ions locate in the same plane with the $Cu^{2+}$ ion, and the other two $Cl^-$ ions situate at the plane of $K^+$ ions, which form the $K_2CuF_4$ local structure. Each crystal structure contains two such $K_2CuF_4$ local structures. In addition, the two $K_2CuF_4$ local structures are separated with large crystal space. Thus, this family of materials has been well recognized as quasi-two-dimensional compounds [50]. Both Polycrystalline and single crystal of $K_2CuF_4$ samples have been prepared in former experiments: polycrystalline samples can be directly prepared from $KCuF_3$, $KZnF_3$ and KF [48]; single crystal $K_2CuF_4$ can be grown by adding a mixture of ground $KCuF_3$ and KF in a 1:1 molar ratio in a conical platinum crucible [47]. In particular, the grown $K_2CuF_4$ single crystal has both high quality and large size [47], which greatly facilitates further experimental characterization. The point group of the crystal structure is tetragonal $D_{4h}^{17}$. For $K_2CuF_4$, the optimized lattice parameters are: a = b = 4.136 Å, c= 13.015 Å. The Cu atoms occupy the 2*a* (0.0, 0.0, 0.0) Wyckoff position and K atoms are located at the Wyckoff positions 4e (0.0, 0.0, 0.356). The F atoms occupy the 4*c* (0.0, 0.5, 0.0) and 4*e* (0.0, 0.0, 0.153) Wyckoff sites, respectively. The crystal structure has been studied by Knox by experiments [47], and the lattice constants were found to be a=b= 4.155Å and c= 12.740 Å, which are well comparable with our optimized ones.

During previous experimental measurements, $K_2CuF_4$ has been characterized as a quasi-two-dimensional ferromagnet with the Curie temperature ($T_c$) being 6.25K [49]. In our calculations, we have also compared the energy between the FM (ferromagnetic) and the AFM (antiferromagnetic) states. We find that the FM state is 99.9 meV lower than the AFM state for a unit cell, also manifesting an FM ground state. In order to determine the direction of spontaneous magnetization, we compare the total energy of the magnetization system along different high symmetry axes based on the GGA+ SOC method. Here the magnetizations along the [100], [110] and [001] directions are taken into account. We find that the energies along the [100] direction and the [001] direction are almost the same, and are a bit lower (0.171 meV) than the [110] direction. The calculated results are consistent with previous experiments, showing the easy *c*-axis magnetization direction, and weak anisotropy [51]. It is worth mentioning that, the weak anisotropy suggests the magnetism in $K_2CuF_4$ is "soft". Therefore the magnetization direction can be easily switched by applying an external magnetic field.

## 3. Results and discussions

Based on the ferromagnetic ground state, we first discuss the electronic band structure of $K_2CuF_4$ in the absence of SOC. The calculation details are displayed in the Supplementary Information. We clearly observe a half-metallic band structure from Fig 2 (a) and (b). It exhibits an insulating property in the spin-up channel with a band gap of 3.7 eV, but the spin-down channel exhibits a metallic band structure. By examining the projected density of states (PDOS), we find that the low-energy states are mainly contributed by the *d* orbitals of Cu atoms, as indicated on the left of Fig.

2(b). From the band structure, the conducting electrons arise from the spin down channel, thus are fully spin polarized.

We pay attention to the band structure in the spin down channel. As shown in Fig. 2(b), we find only two bands appear near the Fermi level. Especially, along the highly symmetric *k*-path $\Gamma$-*X*, these bands exhibit linear dispersions and cross with each other, and form a type-II Weyl point near the Fermi level. For the $\Gamma$-*X* path, the point group is $C_{2v}$. Our symmetry analysis finds the two bands belong to different irreducible representations ($\Gamma_1$ and $\Gamma_2$ respectively), thus the band crossing is unavoidable. After carefully scanning band structure in the entire BZ, we find that the two bands do not cross at an isolated nodal point, but can from a pair of nodal lines in the (110) plane (see the 3D plotting of band structure in Fig 3(a)]. As shown in the enlarged band structure in Fig 3(b), we clearly find that all band crossings at the nodal lines are type-II, indicating the truth of the type-II signature on the whole nodal line. Thus, the nodal lines in $K_2CuF_4$ are type-II. These band crossings have opposite eigenvalues ±1 of mirror reflection symmetry operation $M_{110}$. These nodal lines are protected by the mirror symmetry ($M_{110}$) corresponding to the (110) plane. The mirror plane $M_{(110)}$ *(-y, -x, z)* ensures that the type-II nodal line is on the (110) plane. The profiles of the nodal lines in the BZ are shown Fig 3(c) and (d). We can find the nodal lines transverse the entire BZ along the (110) plane. Such traversing type-II nodal line is also proposed in interpenetrated graphene network (IGN) [52] and $K_4P_3$ [27]. However, the situation in $K_2CuF_4$ is different with these examples, because the type-II nodal line proposed here is completely spin-polarized.

It is well known that, the drumhead surface states are the hallmark of nodal lines. In particular, these nodal lines are fully spin-polarized, and we expect to observe fully spin-polarized surface states. In order to better understand the surface signature of the traversing type-II nodal line in half-metal $K_2CuF_4$, we construct a tight-binding model through the Wannier function. We simplify the TB model by using the K-*s*, Cu-*d* and F-*p* orbitals, which are the main contribution to the low-energy bands. As shown in Fig. 3(e), the band structure from the Wannier function is well consistent with the DFT results. The projected surface states are shown in Fig. 3(f). Obviously, we have observed clear drumhead surface states emanating from the type-II nodal line. To be noted, arising from the fully spin polarized nodal lines, the drumhead surface states here are also fully spin polarized. The clear fully spin polarized drumhead surface states are very helpful for the experimental detection and the practical application from their special transport properties.

Here, we discuss the impact of SOC on the electronic band structure. When SOC exists, symmetry depends on the direction of magnetization. We first consider the spontaneous magnetization direction of the system. As mentioned above, magnetization along the [100] and [001] directions have the lowest energy. But the energy difference among different directions is very small, showing very weak anisotropy. The point group will reduce from $D_{4h}$ to $C_{4h}$ when the magnetization direction is applied along the [001] direction. The mirror symmetry $M_{110}$ is broken, and the nodal lines would be gapped. However, it is noticed that, the system still maintains the $I$, $C_{4z}$, $TC_2^z$, $TC_2^{110}$ symmetry. Especially, the $TC_2^{110}$ symmetry can protect the band crossings in the $\Gamma$-$X$ and $Z$-$Y_1$ paths. This scenario has been verified by our DFT calculations. As shown in the Fig. 4(a), the bands show no gap in the two paths, which are characterized with type-II Weyl nodes. In addition, we have also examined the size of SOC gaps on other parts of the nodal line. Figure 4(b) show the SOC gap at different $k_z$, where $k_z = 0$ and $k_z = 0.5$ represent the $\Gamma$-$X$ and $Z$-$Y_1$ path [see Fig. 3(c)], respectively. We can find that the SOC gap along the nodal line is less than 1.1 meV.

If the magnetization direction is applied along the [110] direction, the mirror symmetry $M_{110}$ preserves even SOC is included. In this case, the group elements of the corresponding magnetic space group $C_{2h}$. The system still remains $I$, $C_2^{110}$. The group element $IC_2^{110}$ of inversion $I$ and rotation $C_2^{110}$ is equivalent to the mirror symmetry in (110) plane. As the result, the type-II nodal lines retain under SOC, as shown in Fig. 4(c). Since both calculations and experiments have shown weak anisotropy in this system, the magnetization direction could be easily changed by applying an external magnetic field. Therefore, the system can exhibit different topological states, such as type-II Weyl nodes and type-II nodal lines, as controlled by the external magnetic field [see Fig. 4(d)].

Before closing, we want to point out that, other members of $X_2YZ_4$ (X=Cs, Rb, Y=Cr, Cu, Z=Cl,) compounds including $Cs_2CrCl_4$, $Rb_2CrCl_4$ and $Rb_2CuF_4$ also show spin-polarized type-II nodal line signature. According to former experiments and our DFT calculations, they are verified to show intrinsic ferromagnetic properties. The spin-resolved band structures are displayed in the Supplementary Information. The optimized (experimental) lattice parameters, the Curie temperature, the total magnetic moment, the size of half-metallic gap, the ratio of spin-polarization, the spin locations of the type-II nodal lines are summarized in Table I. In $X_2YZ_4$ compounds, $K_2CuF_4$, $Cs_2CrCl_4$, $Rb_2CrCl_4$ and $Rb_2CuF_4$, all exhibit the half-metallic band structure with the 100% spin-polarization, where the type-II nodal lines either locate in the spin up or spin down channel [see Table I]. To be noted, the type-II nodal lines in $X_2YZ_4$ compounds are different with those proposed in nonmagnetic system such as $K_4P_3$ [27], $Mg_3Bi_2$ [33,34] and TiBX (X = Cl, Br, I) monolayers [35]. The spin polarization in $X_2YZ_4$ compounds allows them with potential for spin manipulation and spintronics applications. Previously, similar topological half metals have also been proposed to show fermions with 100% polarization, such as $Co_3Sn_2S_2$ [53], monolayer $PtCl_3$ [54], quasi-one-dimensional materials $X_2RhF_6$ (X = K, Rb, Cs) for Dirac/Weyl half metals [55]; $\beta$-$V_2PO_5$ [56], $Li_3(FeO_3)_2$ [57], MnN monolayer [58], CrN monolayer [59], $K_2N$ monolayer [60] for nodal line half metals; some spinel materials for chain-like half metals [61-63]. In addition, massless fermions have also been proposed in other spin-polarized systems such as magnetic semiconductors [64], spin-gapless semiconductors [65,66]. The exploration of spin-polarized fermions will be beneficial to promote the development of spintronic devices.

## 4. Conclusion

In conclusion, we have completely investigated the magnetic configuration, electronic structure, and topological signatures of $K_2CuF_4$ ferromagnet in the family of quasi-two-dimensional compounds $X_2YZ_4$ (X = K, Cs, Rb Y = Cu, Cr, Z = Cl, F). Our results show a FM magnetic ground state with weak anisotropy, which are nicely consistent with existing experimental results. Under the ground state, we find the material show half metal band structure, where only the bands in the spin down channel present near the Fermi level. The band crossing in the spin down channel produces a pair of type-II nodal lines near the Fermi level, locating in the in the (110) plane, protected by the $M_{110}$ symmetry. The nodal lines show clear drumhead surface states, which are also completely spin-polarized. When SOC is taken into account, the material can either show type-II Weyl nodes and type-II nodal lines, depending on the magnetization direction. Our results show that this family of materials is ideal examples to investigate the spin-polarized type-II nodal line fermions and Weyl states without time reversal symmetry.

**Declaration of interests**

The authors declare that they have no known competing financial interests or personal relationships that could have appeared to influence the work reported in this article.

**Acknowledgements**

This work is supported by National Natural Science Foundation of China (Grants No. 11904074). The work is funded by Science and Technology Project of Hebei Education Department, the Nature Science Foundation of Hebei Province (Nos. A2019202222 and A2019202107), the Overseas Scientists Sponsorship Program by Hebei Province (C20200319). The work is also supported the State Key Laboratory of Reliability and Intelligence of Electrical Equipment (No. EERI_PI202000), Hebei University of Technology. One of the authors (X.M. Zhang) acknowledges the financial support from Young Elite Scientists Sponsorship Program by Tianjin.

investigation of the temperature dependence of long-wavelength spin waves in ferromagnetic $Rb_2CrCl_4$, J. Phys. C: Solid State Phys.14 (1981) 5327.

[69] L.C. Gupta, R.Vijayaraghavan, S.D. Damle, U.R.K. Rao, L.D. Khoi, P. Veillet, Magnetic resonance studies in $Rb_2CuF_4$, J. Magn. Reson. 17 (1969) 41-45.

[70] S. Sasaki, N. Narita, I. Yamada, Preparation and Magnetic Susceptibility of $Cs_2CuF_4$ and $Rb_2CuF_4$, J. Phys. Soc. Japan 64 (1995) 2701-2702.

**Table:**

Table 1 The optimized (experimental) lattice parameters, the Curie temperature ($T_c$), the total magnetic moment, the size of half metallic gap, the ratio of spin-polarization, and the spin locations of the type-II nodal lines in $X_2YZ_4$ compounds. The experimental dates are picked from literatures [47]-[49], [67]-[70].

| Compounds | Optimized lattice parameter (Å) | Experimental lattice parameter (Å) | Curie temperature (K) | Total magnetic moment ($\mu_B$) | Size of half metallic gap (eV) | Spin-polarization | Spin locations of the type-II nodal lines |
|---|---|---|---|---|---|---|---|
| $K_2CuF_4$ | a=b=4.137 c=13.015 | a=b=4.155 c=12.740 | 6.25 | 1 | 3.7 | 100% | down |
| $Cs_2CrCl_4$ | a=b=5.167 c=16.903 | a=b=5.163 c=16.379 | 58 | 4 | 4.6 | 100% | up |
| $Rb_2CrCl_4$ | a=b=5.080 c=16.272 | a=b=5.086 c=15.715 | 52.4 | 4 | 4.4 | 100% | up |
| $Rb_2CuF_4$ | a=b=4.246 c=13.607 | - | 6.05 | 1 | 3.4 | 100% | down |

**Figures and captions:**

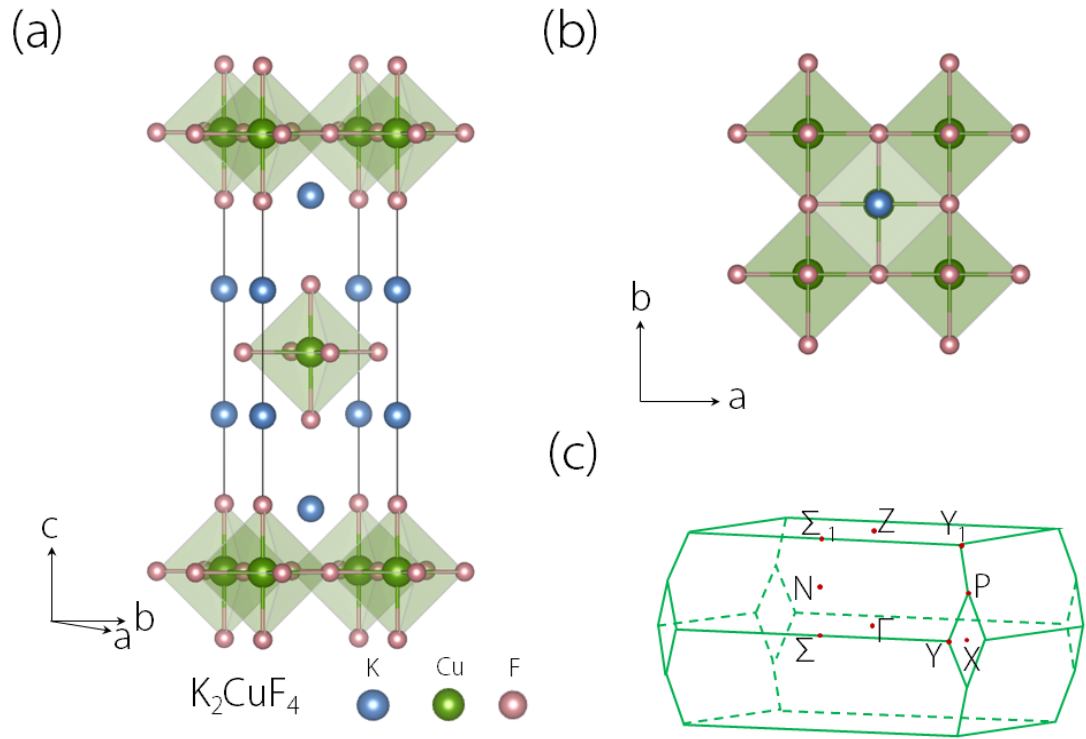

Figure 1 (a) The crystal structure of $K_2CuF_4$ compound. (b) The top view of the crystal structure. (c) The bulk Brillouin zone of $K_2CuF_4$ compound.

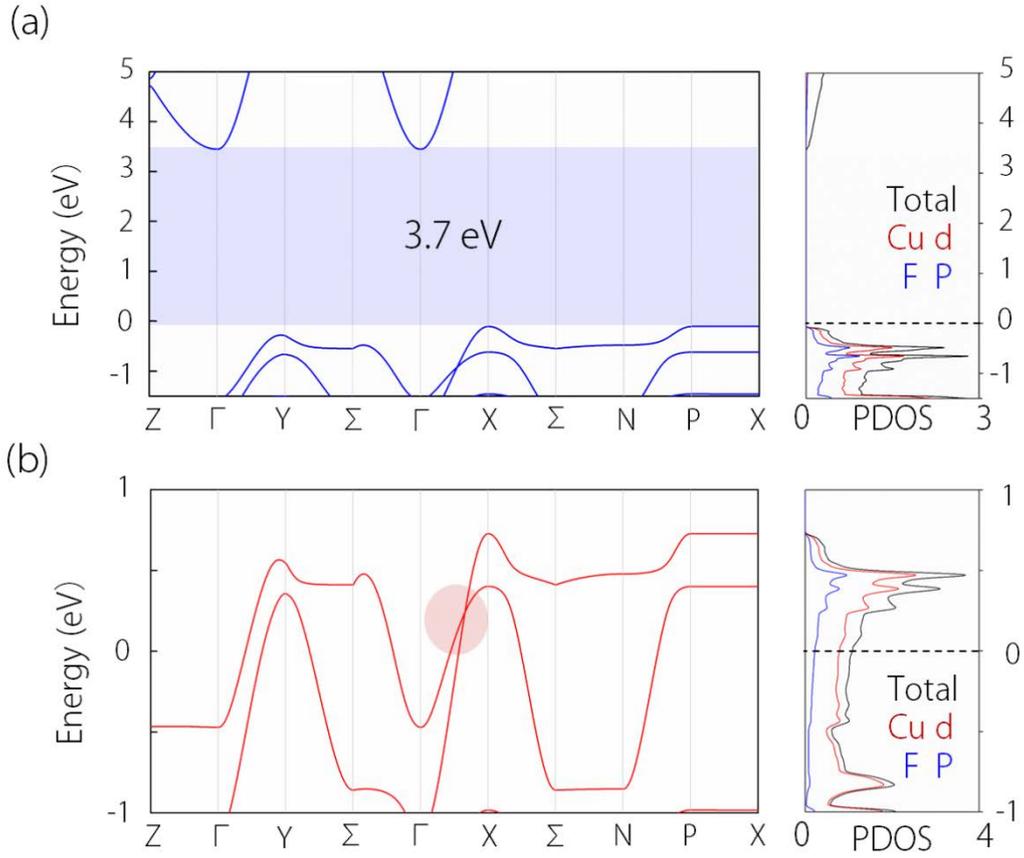

Figure 2 Electronic band structure and the projected density of states of $K_2CuF_4$ compound in (a) spin up channel and (b) spin down channel. In (a), the size of band gap is labeled. In (b), the type-II band crossing is circled by shadow color.

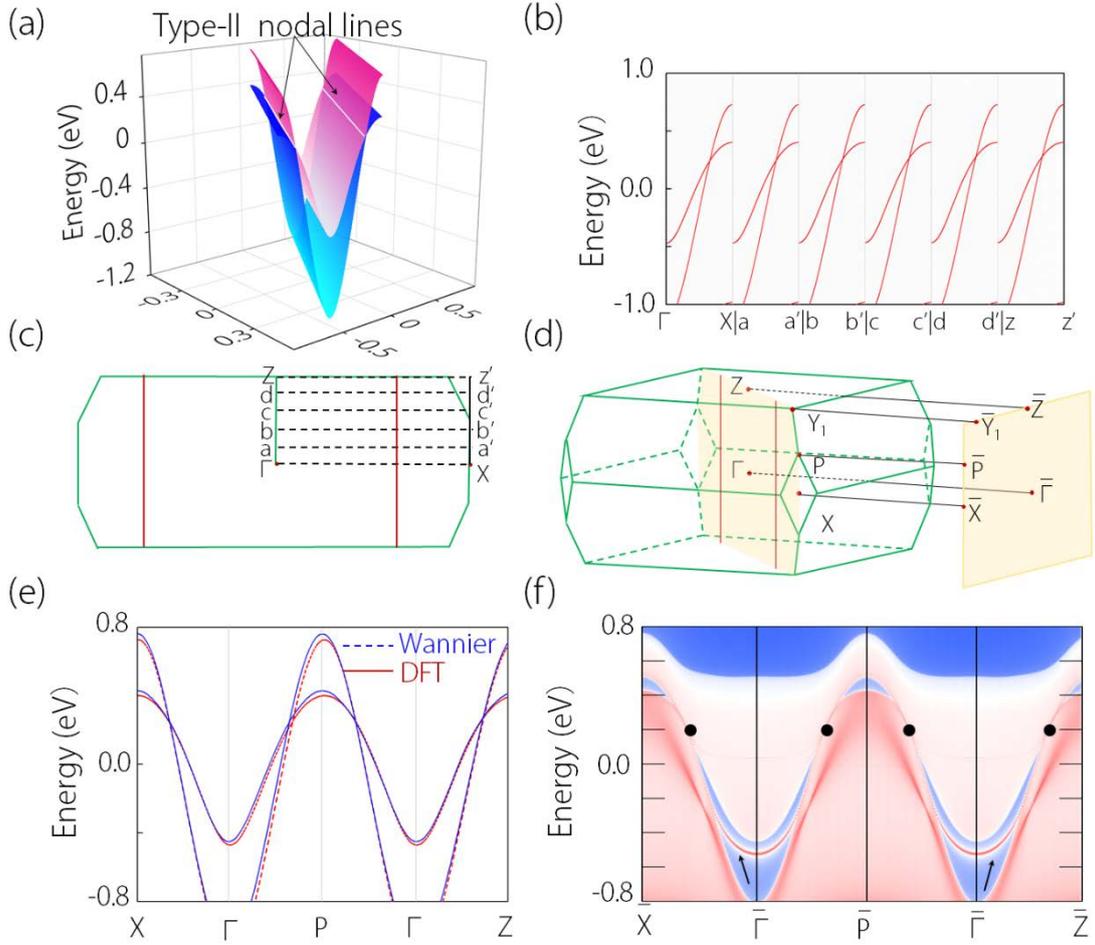

Figure 3 (a) The 3D plot of band dispersions of the type-II nodal lines. The nodal lines are indicated by the white lines. (b)The enlarged electronic band structure along the $\Gamma$-X, a-a´, b-b´, c-c´, d-d´, z-z´ paths in the [110] plane. (c) Schematic diagram of the [110] plane in Brillouin. The k-paths $\Gamma$-X, a-a´, b-b´, c-c´, d-d´, z-z´ are shown in the figure. (d) The shape of type-II nodal line in the Brillouin zone and the corresponding (110) surface Brillouin zone. In (c) and (d), the nodal lines are indicated by the red lines. (e) The comparison of band structure from the Wannier model and DFT. (f) The [110] surface band structure with the drumhead surface states pointed by the black arrows.

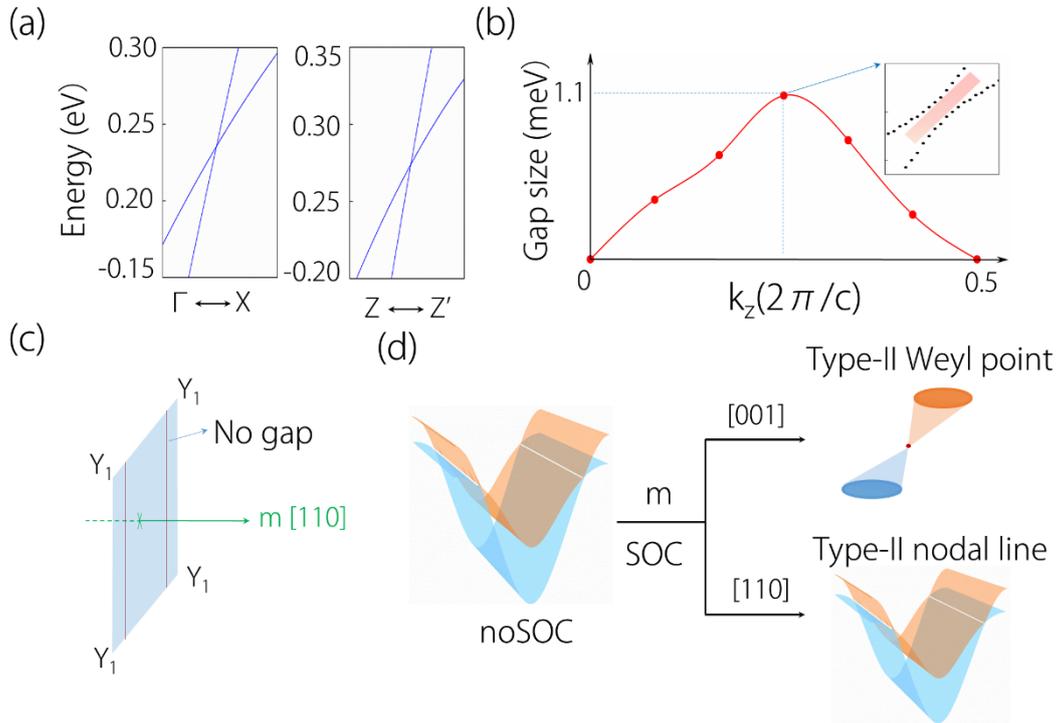

Figure 4(a) The enlarged band structure along the Γ-X, z-z´ paths with SOC under the [001] magnetization direction. (b) The size of SOC gap changes along the $k_z$ direction corresponding to Fig. 3(c). (c) The presence of type-II nodal lines in (110) plane with SOC along the [110] direction. (d) Schematic diagram of different topological phases realized by applying different magnetization directions.